# Recommendations for Emerging Air Taxi Network Operations based on Online Review Analysis of Helicopter Services

Suchithra Rajendran[a,b,*] and Emily Pagel[a]

[a]Department of Industrial and Manufacturing Systems Engineering, University of Missouri Columbia, MO 65211, USA
[b]Department of Marketing, University of Missouri Columbia, MO 65211, USA

## Abstract

The effects of traffic congestion are adverse, primarily including air pollution, commuter stress, and an increase in vehicle operating costs and accidents on the road. In efforts to alleviate these problems in metropolitan cities, logistics companies plan to introduce a new method of everyday commute called air taxis. These are electric-powered vehicles that are expected to operate in the forthcoming years by international transportation companies like Airbus, Uber, and Kitty Hawk. Since these flying taxis are emerging mode of transportation, it is necessary to provide recommendations for the initial design, implementation, and operation. This study proposes managerial insights for these upcoming services by analyzing online customer reviews and conducting an internal assessment of helicopter operations. Helicopters are similar to air taxis in regards to their operations, and therefore, customer reviews pertaining to the former can enable us to obtain insights into the strengths and weaknesses of the short-distance aviation service, in general. A four-stage sequential approach is used in this research, wherein the online reviews are mined in *Stage 1*, analyzed using the bigram and trigram models in *Stage 2*, 7S internal assessment is conducted for helicopter services in *Stage 3*, and managerial recommendations for air taxis are proposed in *Stage 4*. The insights obtained in this paper could assist any air taxi companies in providing better customer service when they venture into the market.

 *Keywords:* Air taxi; Emerging technology; Helicopter services; Online customer reviews; Text analytics; Four-stage sequential approach.

## 1. Introduction

Current traffic conditions in densely populated cities have encouraged new public transit movements. These congestions lead to accidents, increased vehicle operating costs, noise and air pollution, and high commuter stress levels (Rajendran and Zack, 2019).  Air taxi is a revolutionary ridesharing service that is expected to combat ground traffic through aviation. Small electric aircraft are anticipated to provide passengers a faster and more convenient commute compared to the current public transportation systems that are already in place.

These air taxis attempt to significantly alleviate traffic congestion in large metropolitan cities, like New York City and Los Angeles, and could perhaps, become a viable alternative to regular ground taxis (Holden and Geol, 2016). For instance, New York City estimates that nearly four million citizens are traveling each day to Manhattan, resulting in traffic gridlock (Partnership for New York City, 2019). These heavy traffic conditions contribute to the loss of jobs, revenues for business along with other sectors, and economic output. The adverse effects exist in other large cities across the globe (such as Tokyo and London) and have caused the government to encourage commuters to utilize subways or to carpool to work actively.

Although sometimes, fees are also being imposed on individuals who travel through the business districts, heavy traffic still continues to rise due to urbanization (Furfaro, 2018).

Air taxi services, if properly implemented, are anticipated to decrease the negative impact of traffic while also offering customers a new method of transportation. Even though there are many presumed benefits as outcomes of the implementation of this aviation service, several challenges and constraints must be addressed. For example, station location decisions must be made considering variables such as space, customer demand, and infrastructure design in neighborhoods within suburban areas (Antcliff et al., 2016; Hawkins, 2018). Additionally, routing and the coordination of multiple air taxis through a private network must be efficiently monitored and synchronized. A reliable control system is essential due to the existence of considerable commute risk when compared to traveling by road (Rajendran and Zack, 2019).

Although helicopters and air taxis are quite different with regards to the noise generated, travel speed, and flight range, they are very similar in terms of their vertical takeoff and landing (VTOL) operations, scheduling and their infrastructure requirements, such as the presence of helipads for VTOL (Datta et al., 2018). Due to the similarities of business operations, in this study, customer reviews of helicopter services are examined and used to create recommendations for air taxis. By analyzing online customer reviews and conducting an internal assessment of similar services, such as helicopters, we believe that we can obtain insights into the strengths and weaknesses of the short-distance aviation service, in general. Even though air taxis are garnered towards everyday commutes, customers will receive similar experiences and, therefore, may have similar preferences and dislikes. By assessing the current customer-perceived positives and negatives of helicopters, air taxi services gain a greater understanding of ways to appease their passengers and become a successful business in large cities.

The importance of analyzing customer reviews (OCRs) has risen dramatically with the expansion of social media websites and has allowed consumers to assess the quality of a product or service through other people's opinions (Srinivas and Rajendran, 2019). With the increasing accessibility of the Internet, online customer interactions and postings are viewed by thousands of potential purchasers every day, so the distribution of positive reviews is crucial. Since OCRs are considered the second most trustworthy source for product information, satisfied customers who share their experience will encourage other buyers, and therefore, increase product sales (Balaji et al., 2016). The analysis of OCRs also supplements companies' knowledge of customers' perception of their products/services. Utilizing the information provided through the OCRs, the perceived quality of a product/service, or its features, can be determined, and recommendations to improve or sustain can then be made (Wang et al., 2016).

This study uses a four-stage sequential approach, wherein the online helicopter reviews are mined in *Stage 1*, analyzed using the bigram and trigram models in *Stage 2*, 7S internal assessment is conducted for helicopter services in *Stage 3*, and managerial recommendations for emerging air taxi services are proposed in *Stage 4*. The managerial insights could assist air taxi companies to provide better customer service when they venture into the market.

The remaining paper is organized as follows. Section 2 reviews the literature pertaining to air taxi and online customer reviews. Section 3 details the four-stage sequential approach, including the text analytics methodology as well as the organizational effectiveness tool used in this study. Section 4 discusses the results of the internal assessment of helicopter services. Section 5 presents the managerial recommendations for air taxis, while Section 6 gives the discussion questions from the implementation standpoint. The conclusion and future work are given in Section 7.

## 2. Literature Review

### 2.1 Review of Air Taxi

Since air taxis are expected to become a viable alternative to ground transportation for millions of commuters traveling long distances every day, a reliable and safe vehicle that can withstand a large demand is crucial. Falck et al. (2018) discussed the essential aircraft features, including battery life and passenger capacity. Further addressing the flying taxi design, Shamiyeh et al. (2018) evaluated different models, especially during takeoff and landing. It was determined that 18-rotor multicopter design had an advantage with respect to noise levels, but higher travel speed was achieved during the testing of fixed-wing cruise flights. Propulsion methods in aircraft, such as battery-electric, hybrid, and electric/hydrogen, are also studied to determine their efficiency, capacity, and noise level (Anderson et al., 2015; Antcliff et al., 2016; Datta et al., 2018). The advantages and downsides of different models should be considered and prioritized based on the air taxi business methodology.

Although numerous articles have studied the air taxi design, very few papers have focussed on the operational aspects of this aviation service. The weekly flow of passengers for a single provider was evaluated through a network formulation created by Lee et al. (2008). Two models were developed within this study: discrete-event and aggregate flow, and their performances were compared to identify the better performing model under different settings. The authors concluded that the aggregate flow model obtained more accurate outputs and is valid for pricing decisions. To further understand the strategic operations, Sun et al. (2018) estimated the total travel time for passengers using different methods of commutes. The competitiveness of the upcoming market versus ground transportation was analyzed for several populated cities in Europe. The authors suggested that short and very lengthy travel distances are most quickly completed by car and aircraft, respectively. They also concluded that air taxi gains time advantage primarily in three European countries: Belgium, Netherlands, and Germany.

### 2.2 Review of OCR Analysis

With the expansion of the Internet, online reviews allow existing clients to interest potential customers and other readers around the world (Duan et al., 2008; Hu et al., 2008; Zhu and Zhang, 2010; Hu et al., 2014). It was found in a study conducted by Cheung and Lee (2012), that over 61% of the people review customer feedback posted online before purchasing a product or utilizing a service. Additionally, the products that received high ratings (4-5 stars) had at least 20% higher "perception to pay rate."

To further analyze online customer reviews, Erkan and Evans (2016) studied the role of electronic word of mouth (eWOM) as a marketing tool, as well as discussed the relationship with social media sites and consumer opinions. Their article aimed to create an abstract model that would identify the motivators of eWOM in efforts to gain a better understanding of the purchasers' decisions. Likewise, other papers have attempted to analyze the correlation between eWOM and sales (Chen et al., 2004; Chevalier and Mayzlin, 2006; Hu et al., 2008; Zhu and Zhang, 2010; De Maeyer, 2012).

### 2.3 Contributions to the Literature

To the best of our knowledge, this research is the first to propose managerial recommendations for air taxi emerging technology based on the online review analysis of helicopter services. Through the analysis of customer feedback, the current weaknesses and strengths of helicopters are established. These insights provide a greater understanding of customer's satisfaction levels and can be utilized to develop efficient business operations for air taxi services. Moreover, this research is one of the pioneers in integrating internal assessment tools, such as 7S, with text analytics. The developed four-stage sequential approach enables us to assess the current customer-perceived positives and negatives of short-distance aviation service, in general. Even though air taxis are garnered towards everyday commutes, customers will receive similar experiences and, therefore, may have similar preferences and dislikes.

## 3. Methodology

Figure 1 overviews the developed four-stage sequential framework used in this study. Stage 1 pertains to the extraction of online reviews (detailed in Section 3.1). Stage 2 involves the identification of positive and negative feedback using bigram and trigram analysis (Sections 3.2 and 3.3). Step 3 focusses on the internal assessment of helicopter reviews using 7S (Section 3.4). Recommendations for air taxi services based on the results are proposed in Stage 4.

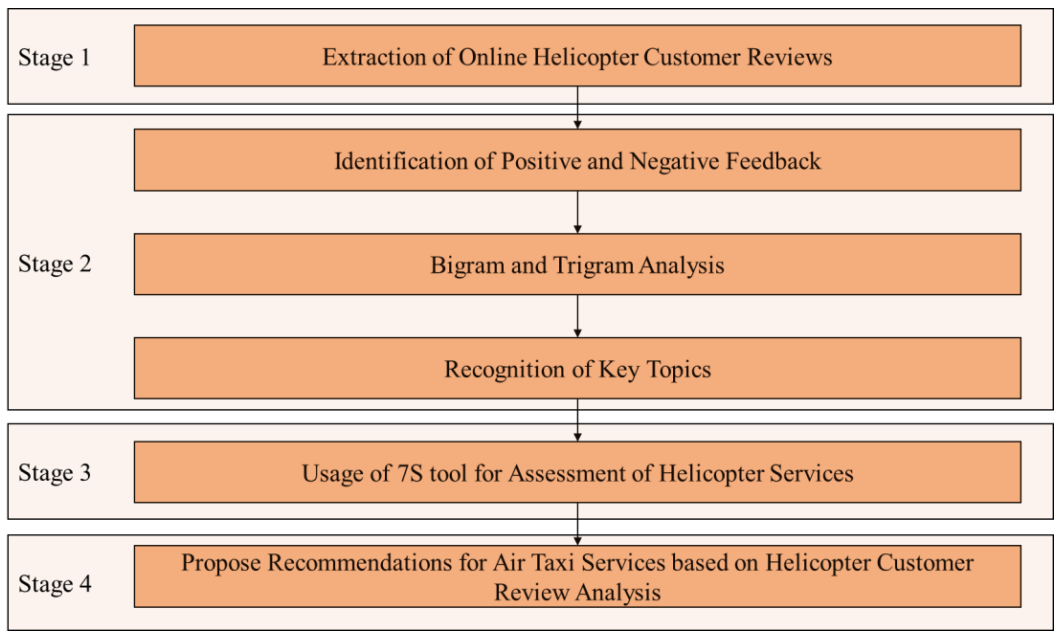

**Figure 1:** Overview of the Proposed Approach

### 3.1 Extraction of Online Customer Reviews and Data Preprocessing

The proposed approach begins with the extraction of publicly available online customer reviews (OCRs) from numerous online sites. For this study, records extracted include service ratings ranging from one to five and the contextual customer reviews. This data is mined to determine most-commonly discussed topics using text analytics. Individual reviews may not be centralized around a single feature of the helicopter service. Thus, contextual feedback is separated into sentences and treated independently. After data extraction, a Python® code is utilized to summarize reviews through the removal of duplicate or missing statements. Furthermore, every sentence is prepared for text analytics through tokenizing, removing unnecessary words, stemming words, and the conversion of all characters to lowercase. These steps are completed through packages available in the Python natural language toolkit (NLTK).

## 3.2 Separation of Reviews based on Star Rating

Upon completion of the review extraction and preprocessing, reviews are separated by the rating score. These ratings range between one and five stars and are dependent on a customer's perception of quality. A rating scale is established to identify the key topics correlated with a positive or negative sentiment. Negative reviews consist of star ratings of 1 and 2, neutral reviews are ratings of 3, and 4-5 ratings are considered positive reviews. The separation of these reviews partition commonly occurring topics between the different categories.

## 3.3 Bigram and Trigram Analysis

The section below discusses the bigram and trigram analysis conducted by Jurafsky and Martin (2014).

Assume that $W_1, W_2, \ldots, W_n$ are a bag of words. The probability of $W_2$ occurring after $W_1$ in a sequence is determined through Equations (1) and (2). Likewise, the probability of the trigram, $W_3$ after $W_2$ and $W_1$, successively is given by Equations (3). Equation (4) derives $P(W_1, W_2, W_3)$ from Equations (2) and (3).

$$P(W_2|W_1) = \frac{P(W_1, W_2)}{P(W_1)} \tag{1}$$

$$P(W_1, W_2) = P(W_2|W_1) \times P(W_1) \tag{2}$$

$$P(W_1, W_2, W_3) = P(W_3|W_1, W_2) \times P(W_1, W_2) \tag{3}$$

$$P(W_1, W_2, W_3) = P(W_1) \times P(W_2|W_1) \times P(W_3|W_1, W_2) \tag{4}$$

Utilizing Equation (4) to $N$-grams, the likelihood of occurrence of a variable $W_N$ in a word series $W_1 \ldots W_{N-1}$ is derived from Equations (5) and (6).

$$P(W_1 \ldots W_N) = P(W_1) \times P(W_2|W_1) \times P(W_3|W_1, W_2) \ldots P(W_N|W_1, W_2 \ldots W_{N-1}) \tag{5}$$

$$P(W_1 \ldots W_N) = \prod_{n=1}^{N} \Pr(W_n|W_1 \ldots W_{n-1}) \tag{6}$$

Through the application of the chain rule of conditional probability to the word series under study, Equation (8) provides the likelihood of the occurrence of any $N$ words.

$$P(W_1 \ldots W_N) = P(W_1) \times P(W_2|W_1) \times P(W_3|W_1^2) \ldots P(W_N|W_1^{n-1}) \tag{7}$$

$$P(W_1 \ldots W_N) = \prod_{i=1}^{N} P(W_N|W_1^{N-1}) \tag{8}$$

Instead, for the bigram model, the Markovian assumption discussed within Equation (9) is studied.

$$P(W_N|W_1^{N-1}) = P(W_N|W_{N-1}) \tag{9}$$

The probability $P(W_N|W_{N-1})$ is approximated using the proportion of the bigram of $W_{N-1}$ and $W_N$ (represented by $\nu(W_{N-1}W_N)$) to the total occurrence of all bigrams under study containing $W_{N-1}$ ($\sum_w \nu(W_{N-1}W)$), as given in Equation (10).

$$P(W_N|W_{N-1}) = \frac{\nu(W_{N-1}W_N)}{\sum_w \nu(W_{N-1}W)} \tag{10}$$

Likewise, the trigram model containing $W_N, W_{N-1}$ and $W_{N-2}$ is derived by constraint (11).

$$P(W_N|W_{N-1}, W_{N-2}) = \frac{\nu(W_{N-2}W_{N-1}W_N)}{\sum_w \nu(W_{N-2}W_{N-1}W)} \tag{11}$$

### 3.4 Introduction to McKinsey 7S Framework

McKinsey's 7S Framework is used to identify the elements within a business model that must be realigned to improve performance and sustainability (Alshaher, 2013). This realignment could include restructuring professional strategies, introducing new processes, and changing staff or leadership. The 7S framework begins with ensuring that shared values are consistent across both hard and soft elements, consisting of structure, strategy, systems, style, staff, and skills (Channon and Caldart, 2015). After establishing these persistent values, both "hard" and "soft" components are cross-referenced to guarantee their support to one another to reach a common objective. If a "soft" metric (shared values, skills, style, and staff) does not reinforce a "hard" component (strategy, structure, and systems), then it must be reconfigured (Hanafizadeh and Ravasan, 2011).

### 4. Results

### 4.1 Case Study

To accommodate customers' preferences almost entirely and propose reputable recommendations for air taxis, customer reviews of helicopter services are analyzed due to their similar business operations. For example, both these transportation methods utilize aircraft to provide passengers rides for short-distance commutes. Their services also face similar obstacles, such as high operating costs and ride scheduling. By considering the current drawback and well-perceived helicopter service features established through online reviews, air taxi services could obtain further insight into aviation customer's likes and dislikes. We mined over 5000 online reviews posted for helicopter services from several social networking sites over the past decade.

### 4.2 Model Summary

Table 1 highlights the critical topics obtained from the bigram and trigram models. While most of the topics, such as promotion, features, staff, reservation, schedule, safety, guided tours, and vehicle maintenance, are viewed positively by customers, negative reviews are obtained for certain topics - e.g., waiting area and site visibility.

**Table 1:** Key Topics Identified using Bigram and Trigram Analyses

| Key Topic | Description |
|---|---|
| Promotion | Exclusive offers, discounts, or coupons given to customers for future services |
| Features | Specific attribute of the aviation service |
| Staff | Individuals employed to serve customers (e.g., pilots, attendants, front-desk staff) |
| Reservation | Booking a future helicopter ride via phone call, online website, or smartphone application |
| Waiting Area | Room/suite designated for customers who are waiting for their helicopter rides |
| Schedule | List of intended future ride details |
| Safety | Unlikelihood of passenger injury, danger, or risk |
| Guided Tour | Aviation ride over historic/well-known area or monument with announcements of specific details |
| Site Visibility | Customers' ability to view through the vehicle windows during their ride |
| Vehicle Maintenance | Service procedure to ensure safe and long-term use of the vehicle |

### 4.3 Analysis of Helicopter Reviews based on 7S Framework

Based on the topics identified, the 7S internal assessment is conducted to understand the "hard" and "soft" components. These elements are then utilized to create recommendations for air taxi services due to operation similarities. Hard features, discussed in Section 3.4, are analyzed first due to their higher impact on businesses' success.

- To begin the internal assessment, the hard element *Strategy* is first studied. Helicopter services frequently offer promotions to increase ride demand and retain returning customers. These discounted services are intended to ensure long-term success and are, therefore a strategic and tactical business decisions. The promotions may be extended to past or future customers and sometimes to large parties depending on aircraft availabilities.
- An additional crucial hard element is *Systems*. For example, helicopter services often handle reservations via online booking websites. These sites are user-friendly and straightforward, which has led to positive customer reviews and can easily be implemented in air taxi service operations. Additional features, like ride payment, are also available on these sites and are considered very convenient by clients.
- The first soft element examined is *Shared Values*. The highest priority for any aviation service is passenger safety, which is emphasized before every helicopter ride, as concluded through online reviews. This value is pertinent for the success of these services, and hence, safety training for all passengers is required before flight. Also, helicopter reviews negatively describe the refund after service cancellation (due to severe weather conditions).

- The second soft element, *Skills*, is also analyzed. Many customers spoke highly of the pilots' flying ability and also commented on captains warning passengers of upcoming turbulence, which led to an increase of comfort. Therefore, to ensure customer safety and comfort level, a high flying safety performance level for all pilots employed by the helicopter service is necessary.
- The last soft element considered in the 7S Framework is *Staff*. In many positive reviews, customers commented on the friendly, helpful staff employed by the aircraft businesses. These included safety trainers, front-desk workers, and pilots. The welcoming environment created by the employees made passengers feel safe and comfortable, which was consistently expressed throughout the reviews. However, sometimes customers (who are particularly obese) feel embarrassed because workers ask for their weight in front of other passengers. Although weight is an essential factor in determining the number of allowable passengers on a helicopter flight, it can be a sensitive topic to customers.

Figure 2 summarizes the results of the internal assessment. The reviews did not provide an evaluation of the "structure" and "systems" components because they are from the management perspective.

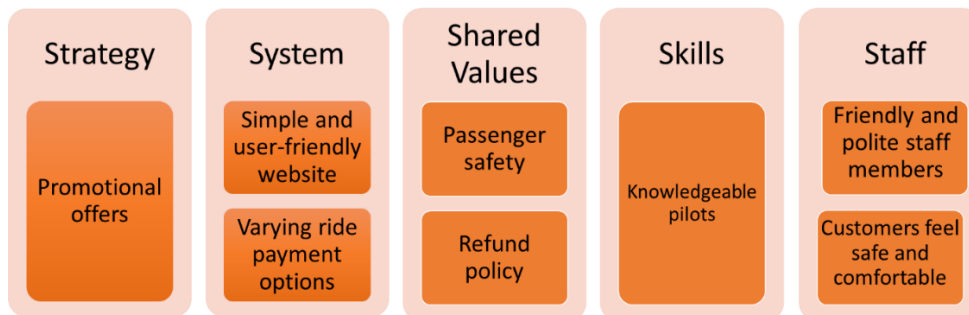

**Figure 2:** Summary of the Internal Assessment of Helicopter Reviews

## 5. Recommendations for Air Taxi Services

Based on the internal assessment of helicopter services using the online positive and negative customer reviews, the following suggestions are proposed for emerging air taxi services.

### 5.1 Promotions

Air taxi services can begin offering coupons and/or free rides to customers on special occasions, such as their birthdays and wedding anniversaries. This incentivizes customers to return annually and may also result in additional marketing for the business. Birthday celebrations are often shared on social media, so customers utilizing their yearly coupons may persuade their followers to avail of the services as well.

Discounts should be offered throughout the year to combat typically high prices per ride. These discounts should be strategically placed during the holidays or seasons with preferable weather conditions. Customers are more likely to purchase a trip during these times, and hence, offering a discount is recommended.

To attract larger parties or corporate events, air taxi services should consider establishing discount rates for passengers paying as a group. Corporations should be offered these discounts at a relatively higher rate, due to their size, to build loyalty. These discounted rates will increase the likelihood of high customer traffic and word-of-mouth marketing.

Establishing a loyalty rewards program encourages customers to return and pay for multiple aviation services. If individuals are incentivized with discounts or free rides for their next visit, they are more likely to return to reap the benefits. Loyalty reward programs can also access personal information, such as birthdays, and track them yearly. The businesses can send coupons via email or creative paper advertisements.

By offering gift cards for air taxis, recipients are more likely to become first-time customers. These gift cards could be sold alongside other transportation services, like Uber or Lyft, at convenience stores, gas stations, grocery stores, or even be purchased online.

To accommodate customers wishing to travel at night, discount fares could be offered after 9 or 10 PM, depending on the marketing strategy. Late-night ride discounts may also attract customers looking to celebrate any occasions at midnight.

For passengers reserving "longer" rides, as determined by the air taxi service, a discount can be applied. This encourages customers to utilize the service for long distances and recommend to others. Price does not need to be linearly increasing with distance.

### 5.2 Built-in Camera

Customers paying for helicopter transportation are often doing so for special occasions, like an anniversary or birthday. Based on this observation from helicopter reviews, we recommend installing and standardizing a high-definition camera in flying taxis so that customers can take photos during their ride. For additional revenue, businesses can sell these photographs as souvenirs (i.e., keychain, photo album, picture frames) to customers after landing.

### 5.3 Staff

The helicopter transportation business standardizes basic aviation training for all employees. If all employees are highly knowledgeable, customers will feel more comfortable and safer before their ride. During the initial stages of operation, any hesitant passengers can choose to be partnered with a highly trained assistant throughout their trip at an additional cost.

### 5.4 Reservations

Air taxi services should offer passengers the ability to reserve specific seats within the vehicle to accommodate preferences. Preferred seats, like seats near the window, could be booked at a higher price. Their preference may depend on the type of travel, whether for everyday commuting or special occasions.

Helicopter reviews negatively describe the refund after service cancellation (due to severe weather conditions). Therefore, for air taxi services, we recommend to

    (a) establish a policy granting a free/discounted rescheduled ride,
    (b) give complete or partial refund immediately, or
    (c) offer alternative transportation if customers utilizing the air taxi for daily commuting, which can consist of train tickets or regular taxis.

These scheme(s) should be clearly displayed on a business' website or shown to customers before reservation is made to gain their trust.

## 5.5 Waiting Area

To combat longer waiting times for air taxis, a waiting area should be created at certain stations with high demand. These areas could have children's activities, televisions, and, if possible, some refreshments, to maximize customer comfort. Standardizing waiting areas will improve a passenger's overall experience and decrease the likelihood of spreading negative eWOM regarding long wait times.

## 5.6 Schedule

To decrease customer's wait time, air taxi service should establish a time buffer in between reservation times. This allows the pilot to have sufficient time to prepare for the next flight while also compensating for any flights running behind schedule, or vehicle maintenance times. If air taxi rides continue to be late after standardizing time buffer, additional measurements, such as the implementation of lean and six sigma techniques, should be taken.

Besides, to alleviate customer dissatisfaction when rides are late, coupons or future ride discounts should be made available.

## 5.7 Safety

Before a customer can successfully reserve a ride, online safety training should be required. By using this policy, the need for training on-site could be eliminated, which may be time-consuming. Once an individual successfully finishes the training, he/she might not have to retake it for future rides. The safety courses can be presented as videos or text to the passenger, depending on their choice. Customers may also choose to receive safety training on-site based on their preference.

Although weight is an essential factor in determining the number of allowable passengers on an aviation service, it can be a sensitive topic to obese customers. By requiring necessary personal information (e.g., height, weight, age) on the phone application or website, the need to question the passenger at the site in front of other travelers is eliminated. This will be a more private and discretionary way to gather this necessary data.

If pilots are aware of upcoming turbulence, they should communicate this in detail with the passengers. When the individuals are prepared for bumps along the ride, they are more likely to feel safe and comfortable on the flight. The pilots should also have an open discourse with all passengers regarding the air taxi's status during the entirety of the ride.

## 5.8 Guided Tours

As an additional service, air taxis should offer guided tours around historical or popular areas within a city. These tours can offer discounted group rates to attract tourists, as well. The pilots should be very knowledgeable of the historical significance of all monuments, cities, and landmarks in the area.

## 5.9 Site Visibility

To ensure that passengers can successfully locate air taxi sites, signboards should be clearly displayed around the stations. It is also recommended that billboards be placed on major highways that contain directional information about the facilities.

## 5.10 Vehicle Maintenance

Reviews praise helicopter service's maintenance and cleanliness. For expensive travel, like air taxis, customers would have higher expectations for ride comfortability. To meet passengers' expectations, it is recommended that air taxis are regularly cleaned and maintained. Customers should also be given a survey upon completion of the trip regarding their ride experience to gather useful data. If passengers express their concern regarding cleanliness, additional measures must be taken.

Figure 3 briefs the key recommendations for air taxi based on the review analysis.

| | |
|---|---|
| **Promotion** | • Promotional offers for special occasions, late night rides, longer duration trips, loyal customers, first time users |
| **Build-in Camera** | • Installing cameras and selling souvenirs |
| **Staff** | • Friendly staff members |
| **Reservations** | • Booking specific seats<br>• Easy refund policy |
| **Waiting Area** | • Introducing waiting area to maximize customer comfort |
| **Schedule** | • Developing efficient scheduling system allowing buffer time between reservations |
| **Safety** | • Proper safety instructions and prior warning of turbulence<br>• Gathering information, such as weight and age, using the app |
| **Guided Tours** | • Attraction of tourist |
| **Site Visibility** | • Proper signboards in the neighborhood |
| **Vehicle Maintenance** | • Customer survey after every ride<br>• Time allowance for cleaning and maintenance |

**Figure 3:** Summary of the Major Air Taxi Recommendations

## 6. Discussion

Although air taxi companies attempt to provide millions of individuals with a more comfortable commute at a competitive cost, certain concerns might arise from potential customers. Numerous papers have focussed on the prospective design and operations of air taxi services (e.g., Holden and Geol, 2016; Johnson et al., 2018; Sun et al., 2018; Rajendran and Zack, 2019); however, the following questions need to be addressed by companies from the implementation standpoint.

The following are a couple of questions and concerns that might prevail among individuals.

- Air taxi services, in efforts to improve accessibility and increase the customer market, could serve passengers with disabilities. Given the rapid customer loading and unloading time, how can service for these customers be accommodated?

- This service may also apply to emergency situations in addition to everyday commuting. Rather than offering this to an entire population, a selected community, like a retirement complex, may be chosen. Could these services be provided to a retirement community so that they can get to the hospital faster? Or could air taxis be used to transfer organs or blood to hospitals and surgery centers?

- To improve the safety and comfort level of passengers, standardized aircraft cleaning and maintenance should be established. How can the successful implementation of cleaning services and maintenance checks be created within a busy ride schedule?

- Due to the small size and weight constraints of aircraft, would there be challenges to commute passengers who travel to and from airports who are more likely to carry heavy pieces of baggage?

- To ensure passenger safety, procedures, and security checks are required by aviation services. However, with scheduling constraints, this expected additional time may pose many problems for air taxis. How is the air taxi service planning to handle standardized safety procedures and security check time?

- Many promotional offers available from helicopter services focus on rates for large groups or events, but do not consider discounts for young passengers. What pricing strategies for children are economically plausible?

## 7. Conclusion

Currently, several logistics companies are exploring the design of air taxis in large metropolitan cities. These electric-powered aircraft vehicles operate similar to helicopters, although the former are in the process of serving millions of everyday commuters in efforts to alleviate traffic congestion and reduce travel time. Due to their correlating business operations, the current strengths and weaknesses of helicopter services are assessed to determine plausible recommendations for air taxi's design, operation, and implementation. The advantages and shortcomings of aviation service are inferred after performing text analysis on helicopter services' online customer reviews (OCRs). These OCRs often contain vital information regarding the customers' perceptions of the service and may impact potential customers' decision to avail of the emerging methods.

This study develops a four-stage sequential approach. To determine insights that may improve the design and business operations of air taxis, OCRs are extracted from multiple social networking sites in Stage 1, and text analytics on OCRs is conducted in Stage 2. The analysis includes positive and negative review separation, bigram and trigram identification, and the recognition of critical topics based on commonly co-occurring words. Upon the completion of text analysis, a 7S internal assessment is conducted for helicopter services in Stage 3, and several managerial recommendations for air taxis are proposed in Stage 4.

The helicopter service OCRs provided many interesting findings that can be used for air taxi recommendations. For example, many customers speak highly of their ride experience, especially when accompanied by a reliable and polite staff or when additional features like a built-in vehicle camera are available. Also, discounted rates could be applicable to large groups, corporate events, or returning passengers to ensure a high customer retention rate. Safety is also a considerable concern for helicopter customers. To compensate for this, mandatory standardized safety training prior to the first air taxi ride is recommended. The proposed recommendation can assist logistic company for efficient network operation.

One of the major limitations of this research is the lack of adequate number of online reviews. With the availability of a large amount of data, better text analytic techniques, such as sentiment analysis and topic modeling tools, can be used for online review analysis, and more valuable recommendations could be provided.

**References**

Alshaher, A. A. F. (2013). The McKinsey 7S model framework for e-learning system readiness assessment. International Journal of Advances in Engineering & Technology, 6(5), 1948.

Anderson, K., Blanchard, S., Cheah, D., Koling, A., & Levitt, D. (2015). City of Oakland Mobility Hub Suitability Analysis Technical Report. (December).

Antcliff, K. R., Goodrich, K., & Moore, M. (2016, March). NASA silicon valley urban VTOL air-taxi study. In On-demand mobility/emerging tech workshop, Arlington (Vol. 7).

Balaji, M. S., Khong, K. W., & Chong, A. Y. L. (2016). Determinants of negative word-of-mouth communication using social networking sites. Information & Management, 53(4), 528-540.

Channon, D. F., & Caldart, A. A. (2015). McKinsey 7S model. Wiley encyclopedia of management, 1-1.

Chen, P. Y., Wu, S. Y., & Yoon, J. (2004). The impact of online recommendations and consumer feedback on sales. ICIS 2004 Proceedings, 58.

Cheung, C. M., & Lee, M. K. (2012). What drives consumers to spread electronic word of mouth in online consumer-opinion platforms. Decision support systems, 53(1), 218-225.

Chevalier, J. A., & Mayzlin, D. (2006). The effect of word of mouth on sales: Online book reviews. Journal of marketing research, 43(3), 345-354.

Datta, A., Elbers, S., Wakayama, S., Alonso, J., Botero, E., Carter, C., . . . Martins, F. (2018). Commercial Intra-City On-Demand Electric-VTOL Status of Technology. Retrieved from https://vtol.org/files/dmfile/TVF.WG2.YR2017draft.pdf

De Maeyer, P. (2012). Impact of online consumer reviews on sales and price strategies: A review and directions for future research. Journal of Product & Brand Management, 21(2), 132-139.

Duan, W., Gu, B., & Whinston, A. B. (2008). The dynamics of online word-of-mouth and product sales—An empirical investigation of the movie industry. Journal of retailing, 84(2), 233-242.

Erkan, I., & Evans, C. (2016). The influence of eWOM in social media on consumers' purchase intentions: An extended approach to information adoption. Computers in Human Behavior, 61, 47-55.

Falck, R. D., Ingraham, D., & Aretskin-Hariton, E. (2018). Multidisciplinary Optimization of Urban-Air-Mobility Class Aircraft Trajectories with Acoustic Constraints. In 2018 AIAA/IEEE Electric Aircraft Technologies Symposium (p. 4985).

Furfaro, Danielle, et al. "Why Driving in NYC Has Somehow Gotten Even Slower." New York Post, New York Post, 16 June 2018, nypost.com/2018/06/15/why-driving-in-nyc-has-somehow-gotten-even-slower/.

Hanafizadeh, P., & Ravasan, A. Z. (2011). A McKinsey 7S model-based framework for ERP readiness assessment. International Journal of Enterprise Information Systems (IJEIS), 7(4), 23-63.

Hawkins, A. J. "Airbus' Autonomous 'Air Taxi' Vahana Completes Its First Test Flight." The

Holden and Goel. "Fast-Forwarding to a Future of On-Demand Urban Air Transportation." 27 October 2016.

Hu, N., Koh, N. S., & Reddy, S. K. (2014). Ratings lead you to the product, reviews help you clinch it? The mediating role of online review sentiments on product sales. Decision support systems, 57, 42-53.

Hu, N., Liu, L., & Zhang, J. J. (2008). Do online reviews affect product sales? The role of reviewer characteristics and temporal effects. Information Technology and management, 9(3), 201-214.

Jensen, P. A., & Bard, J. F. (2003). Operations research models and methods (Vol. 1). John Wiley & Sons Incorporated.

Johnson, W., Silva, C., & Solis, E. (2018). Concept Vehicles for VTOL Air Taxi Operations.

Lee, D. W., Bass, E. J., Patek, S. D., & Boyd, J. A. (2008). A traffic engineering model for air taxi services. Transportation Research Part E: Logistics and Transportation Review, 44(6), 1139-1161.

Rajendran, S., & Zack, J. (2019). Insights on strategic air taxi network infrastructure locations using an iterative constrained clustering approach. Transportation Research Part E: Logistics and Transportation Review, 128, 470-505.

Shamiyeh, M., Bijewitz, J., & Hornung, M. (2017). A Review of Recent Personal Air Vehicle Concepts. In Aerospace europe 6th ceas conference (913, pp. 1–18).

Srinivas, S., & Rajendran, S. (2019). Topic-based knowledge mining of online student reviews for strategic planning in universities. Computers & Industrial Engineering, 128(1), 974-984.

Sun, X., Wandelt, S., & Stumpf, E. (2018). Competitiveness of on-demand air taxis regarding door-to-door travel time: A race through Europe. Transportation Research Part E: Logistics and Transportation Review, 119, 1-18.

Tan, P. N., Steinbach, M., & Kumar, V. (2006). Cluster analysis: basic concepts and algorithms. Introduction to data mining, 8, 487-568.

Verge, The Verge, 1 Feb. 2018, https://www.theverge.com/2018/2/1/16961688/airbus-vahana-evtol-first-test-flight

Wagstaff, K., Cardie, C., Rogers, S., & Schrödl, S. (2001, June). Constrained k-means clustering with background knowledge. In ICML (Vol. 1, pp. 577-584).

Wang, T., Yeh, R. K. J., Chen, C., & Tsydpov, Z. (2016). What drives electronic word-of-mouth on social networking sites? Perspectives of social capital and self-determination. Telematics and Informatics, 33(4), 1034-1047.

Warwick, G. (2018). New Zealand welcomes flight tests of Kitty Hawk's eVTOL air taxi: full-scale prototypes of Cora air taxi in flight testing; transitional eVTOL combines rotors for vertical flight with wings for efficient forward flight. Aviation Week & Space Technology.

Zhu, F., & Zhang, X. (2010). Impact of online consumer reviews on sales: The moderating role of product and consumer characteristics. Journal of marketing, 74(2), 133-148.